\documentclass[prl, superscriptaddress,twocolumn,floatfix,showpacs, reprint]{revtex4-1}

\usepackage{graphicx}
\usepackage{amsmath, amssymb}
\usepackage[usenames]{color}
\graphicspath{{Figs/}}

\begin{document}
\title{A rheological signature of frictional interactions in shear thickening suspensions} 
\author{John R. 
Royer}

\affiliation{Materials Science and Engineering Division, National Institute of Standards and Technology, 
Gaithersburg, MD 20899}

\author{Daniel L. Blair}

\affiliation{Department of Physics and Institute for Soft Matter Synthesis and Metrology, Georgetown University, Washington, DC 20057}

\author{Steven D. Hudson}
\affiliation{Materials Science and Engineering Division, National Institute of Standards and Technology, 
Gaithersburg, MD 20899}

\date{ \today} 

\begin{abstract}

Colloidal shear thickening presents a significant challenge because the macroscopic rheology becomes increasingly controlled by the microscopic details of short ranged particle interactions in the shear thickening regime. Our measurements here of the first normal stress difference over a wide range of particle volume fraction elucidate the relative contributions from hydrodynamic lubrication and frictional contact forces, which have been debated. At moderate volume fractions we find $N_1<0$, consistent with hydrodynamic models, however at higher volume fractions and shear stresses these models break down and we instead observe dilation ($N_1>0$), indicating frictional contact networks. Remarkably, there is no signature of this transition in the viscosity, instead this change in the sign of $N_1$ occurs while the shear thickening remains continuous. These results suggest a scenario where shear thickening is driven primarily by the formation of frictional contacts, with hydrodynamic forces playing a supporting role at lower concentrations. Motivated by this picture, we introduce a simple model which combines these frictional and hydrodynamic contributions and accurately fits the measured viscosity over a wide range of particle volume fraction and shear stress. 

\end{abstract}

\maketitle

There is mounting evidence from recent experiments \cite{Guy:2015aa,Lin:2015aa} and simulations \cite{Mari:2015ab} suggesting that contact friction plays a dominant role in colloidal shear thickening, however this assertion is controversial because of contrary evidence. While friction-based models and simulations capture the viscosity increase observed in experiments, other experimental signatures, particularly the stress anisotropy, are at odds with expectations for frictional interactions \cite{Gurnon:2015aa}. 

Shear thickening, where a suspension's viscosity $\eta = \sigma/\dot{\gamma}$ increases with increasing shear stress $\sigma$ (or shear rate $\dot{\gamma}$), is important in a wide array of industrial processes and applications, either something to be avoided or a desired, engineered property \cite{Barnes:1989aa, Brown:2014aa, Mewis:2011aa}. Shear thickening is observed in both granular suspensions, where the particle diameter $d$ is generally $d\gtrsim 10$ $\mu$m, and colloidal suspensions, where $d\lesssim 10$ $\mu$m. In granular suspensions, the evidence that friction drives shear thickening is well established \cite{Fall:2008aa, Brown:2012qf, Fernandez:2013aa, Heussinger:2013aa, Seto:2013aa, Xu:2014ly, Mari:2014aa, Wyart:2014ve, Ness:2016aa} but in colloidal suspensions shear thickening is instead commonly attributed to diverging hydrodynamic lubrication forces, which lock particles together in correlated `hydroclusters'  \cite{Melrose:1995aa, Phung:1996aa,  FOSS:2000aa, Wagner:2009aa, Cheng:2011aa}.

A key difference between friction and lubrication forces lies in the stress anisotropy generated by these two types of interactions. This difference is captured by the first normal stress difference $N_1 \equiv \sigma_{xx} -\sigma_{zz}$, where $\sigma_{ij}$ is the stress tensor for a shear flow in the $x$ direction with a gradient along $z$. Simulations based on hydrodynamic interactions show that shear-induced distortions of the suspension microstructure and short ranged lubrication forces drive $N_1<0$ \cite{Phung:1996aa, FOSS:2000aa, Bergenholtz:2002aa, Mewis:2011aa}. Including repulsive interactions or elastic particle deformations to these hydrodynamic models does not change the sign of $N_1$ \cite{Melrose:2004ab,Melrose:2004aa, Jamali:2015aa}, and  $N_1$ is predicted to become increasingly negative as the particle concentration increases. In contrast, dilatancy ($N_1>0$) is a well known feature of dense, frictional granular materials \cite{Reynolds:1885aa, Brown:2012qf}, reflecting the anisotropic nature of the force chain network \cite{Cates:1998aa}.  

While proposed friction-based models for shear thickening do not make explicit predictions for $N_1$, at sufficiently high volume fractions one expects frictional contact networks to lead to dilation ($N_1>0$), as in the granular case. Only a handful of experiments measure $N_1$ in shear thickening colloids, though most report $N_1 <0$ \cite{Laun:1994aa, Lee:2006aa, Cwalina:2014aa, Gurnon:2015aa}, consistent with lubrication forces; the lone exception is a study using roughened particles \cite{Lootens:2005aa}. Recent experimental evidence for friction-driven colloidal shear thickening focuses on the viscosity alone, either comparing viscosity profiles to friction-based models \cite{Guy:2015aa} or using shear-reversal techniques to separate contributions from hydrodynamic and contact forces \cite{Lin:2015aa}, and thus these experiments do not address this discrepancy in the sign of $N_1$.

%{\color{red} Dilatancy ($N_1>0$) is a well known feature of frictional granular materials \cite{Reynolds:1885aa, Brown:2012qf}, reflecting the anisotropic nature of the force chain network \cite{Cates:1998aa}.  Hydrodynamic lubrication interactions instead distort the particle pair distribution under shear to generate an excess number of contacts in extension, so that $N_1<0$ \cite{Phung:1996aa, FOSS:2000aa, Bergenholtz:2002aa}. Repulsive forces can be added to hydrodynamic models to change the viscosity profile but these forces do not change the sign of $N_1$ \cite{Melrose:2004ab,Melrose:2004aa, Jamali:2015aa}, making $N_1$ a unique diagnostic of frictional interactions. While only a handful of experiments measure $N_1$ in shear thickening colloids, most report $N_1 <0$ \cite{Laun:1994aa, Lee:2006aa, Cwalina:2014aa, Gurnon:2015aa}, consistent with lubrication forces;  the lone exception is a study using roughened particles \cite{Lootens:2005aa}. Recent evidence for friction-driven colloidal shear thickening focuses on the viscosity alone, either comparing viscosity profiles to friction-based models \cite{Guy:2015aa} or using shear-reversal techniques to separate contributions from hydrodynamic and contact forces \cite{Lin:2015aa}, and thus these experiments do not address this discrepancy in the sign of $N_1$. }

In this Letter we address this disagreement between friction-based models and experiments. Detailing the behavior of both the viscosity $\eta(\sigma, \phi)$ and $N_1(\sigma, \phi)$ over a wide range of shear stresses and volume fractions in colloidal silica spheres exhibiting continuous shear thickening, we show that negative contributions to $N_1$ from lubrication forces can mask positive frictional contributions at moderate volume fractions, but at sufficiently high volume fractions and stresses, frictional interactions become dominant and $N_1$ transitions from negative to positive. This  highlights the need to include {\it both} lubrication and friction to fully describe shear thickening at moderate volume fractions, suggesting possible modifications to purely friction-based models for shear thickening.

\begin{figure*}[t]
\center{\includegraphics[width=14.65cm]{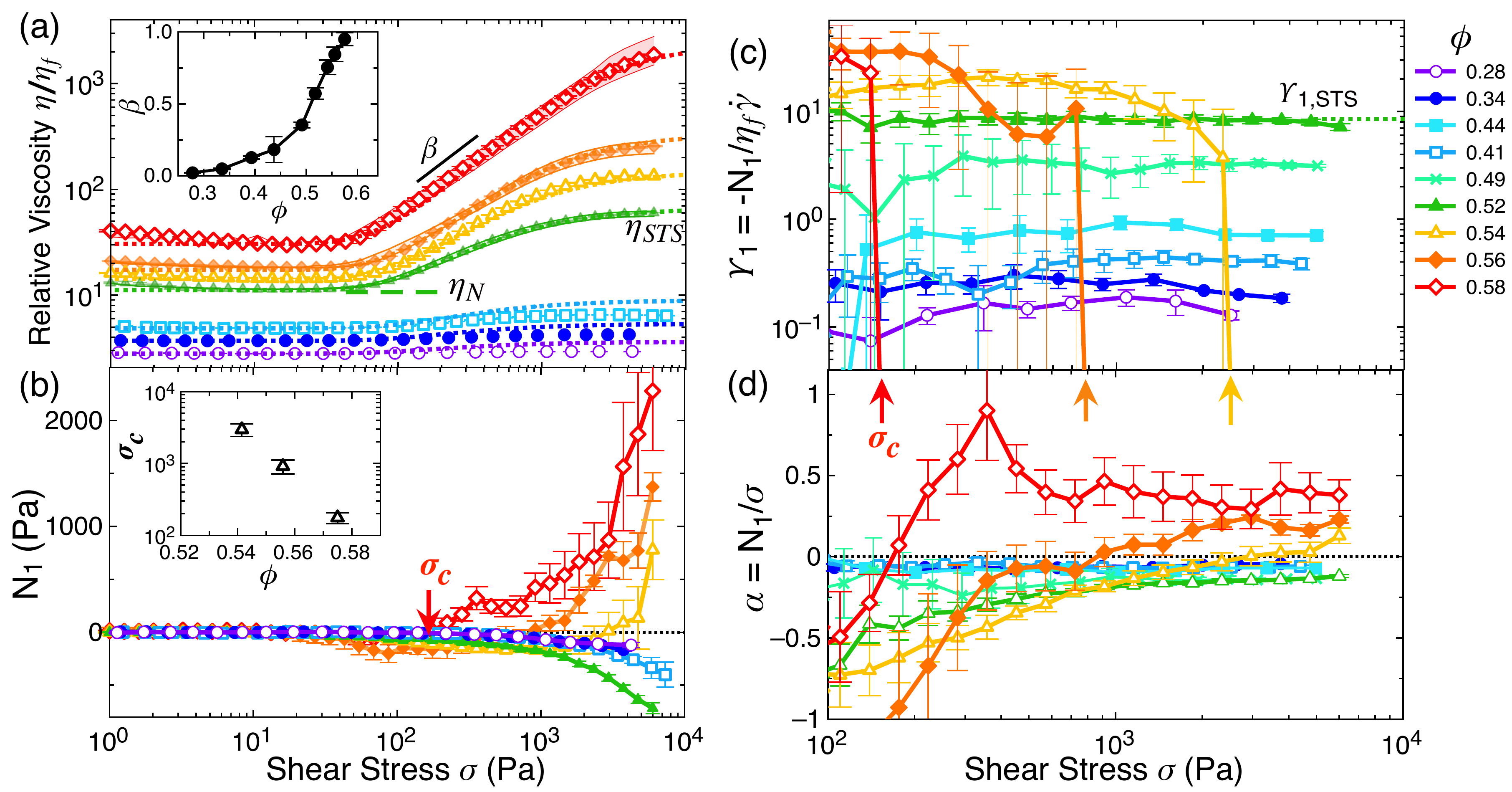}}
\caption{(Color) Transition to dilation in shear thickening suspensions. (a) Relative viscosity $\eta_r (\sigma, \phi)$. Dotted lines: fits to Eq. (\ref{eq:WC}) Inset: shear thickening exponent $\beta(\phi)$. (b) First normal stress difference $N_1(\sigma, \phi)$. Inset: crossover stress $\sigma_c(\phi)$ where $N_1$ crosses zero.  (c) First normal stress difference coefficient $\varUpsilon_1 \equiv -N_1/\eta_f \dot{\gamma}$. See the Supplemental Material for the $N_1<0$ results on an expanded scale \cite{sup_note1}.   (d) Stress ratio $\alpha=N_1/\sigma$. Error bars reflect the standard deviation from multiple up and down stress sweeps. Stress sweeps are conducted at several fixed temperatures: $T= 1$ $^{\circ}\textrm{C}$, 10 $^{\circ}\textrm{C}$,  21 $^{\circ}\textrm{C}$, and 35 $^{\circ}\textrm{C}$. Shaded regions in (a) show the range of $\eta_r$ from $T=1^{\circ} \textrm{C}$ (upper bound) to $T=21 ^{\circ} \textrm{C}$ (lower bound) for $\phi \geq 0.52$, all other quantities are independent of $T$.}
\label{fig:rel_visc_n1}
\end{figure*} 

Here we work with unmodified $d=1.54$\,$\mu$m silica spheres (Bang Laboratories, Inc. \cite{nist_prod}) suspended in a glycerol/water mixture (92\% glycerol mass fraction). A small amount of salt  is added to screen electrostatic interactions ([NaCl]$=0.001$ mole/L), so that the Debye screening length $\kappa^{-1} = 7$\,nm is small compared to the particle size. We prepare samples with volume fractions $0.28 \leq \phi \leq 0.58$ from a concentrated stock suspension with $\phi_{stk} =0.58$, which we determine from confocal imaging and particle locating in samples that have been index matched and diluted by a known ratio. Our relative uncertainty in $\phi$ due to uncertainty in $d$ and particle locating errors is approximately 3\%, e.g. $\phi_{stk} = 0.58$ $\pm$ 0.02.

Rheology is performed under steady shear using an Anton-Paar MCR302 with a $R=12.5$\,mm radius cone-plate tool. In this geometry, $N_1$ can be measured from the axial force $N_1 = 2 F_z/\pi R^2$.  In order to access large shear stresses at all volume fractions over limited range of shear rates, we perform stress sweeps at fixed temperatures between $T=1$\,$^\circ$C and $T=35$\,$^\circ$C to adjust the viscosity of the suspending fluid between $\eta_f = 1.7$~Pa\,s to $\eta_f = 0.107$~Pa\,s. For $\phi\leq 0.52$, changing $\eta_f$ has no impact on either the relative viscosity $\eta_r(\sigma) = \eta(\sigma)/\eta_f$ or $N_1(\sigma)$, and thus on the onset stress for shear thickening.  At higher $\phi$ there is a slight increase in the shear thickening with increasing $\eta_f$, though this variation is small compared to the variation between samples at different $\phi$. At all $\phi$ both $\eta_r(\sigma)$ and $N_1(\sigma)$ are reversible, with no observable hysteresis in repeated up and down stress sweeps. Similarly, we do not observe time dependence at fixed $\sigma$, indicating that the flow curves in Figure~\ref{fig:rel_visc_n1} reflect steady-state suspension properties.  See the Supplemental Material for additional details \cite{sup_note1}.  

\begin{figure}
\center{\includegraphics[width=7cm]{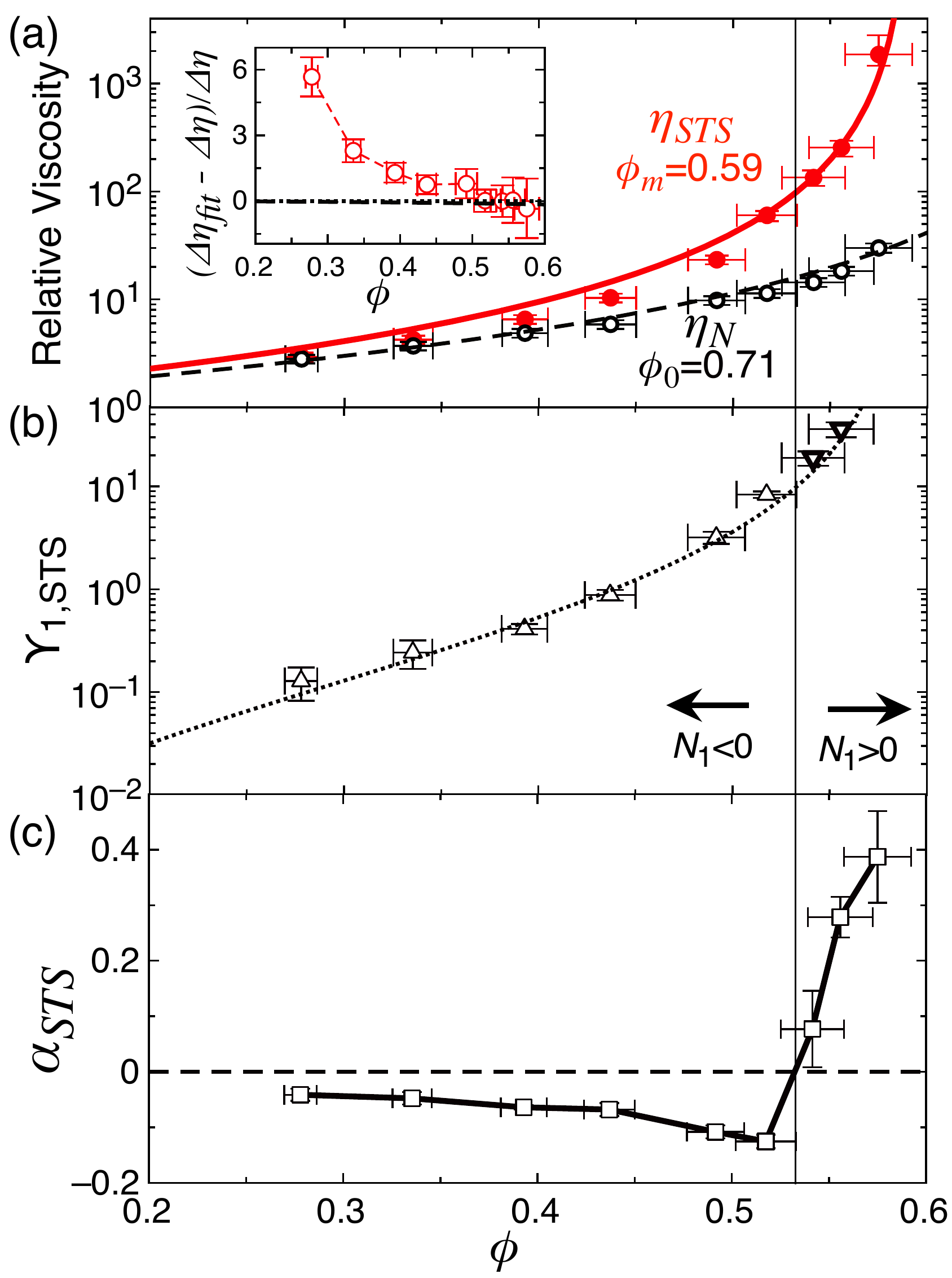}}
\caption{(Color online) Limiting behavior of $\eta_r$ and $N_1$. (a) Newtonian viscosity $\eta_N$ (open circles) and the shear thickened plateau $\eta_{STS}$ (solid circles). Lines show fits  $\eta_N=(1-\phi/\phi_0)^{-2}$ (dashed line) and $\eta_{STS}=(1-\phi/\phi_m)^{-2}$ (solid line). Inset: difference between measured $\Delta \eta \equiv \eta_{STS}- \eta_N$ and fitted expressions. (b)  $\varUpsilon_{1,STS}(\phi)$. Triangles: values where $N_1<0$. Upside-down triangles: plateau values below $\sigma_c$. Dotted line: fit to Eq.~(\ref{eq:MB}). (c) Stress ratio $\alpha_{STS}$. Uncertainties in $\phi$ are 3\% as noted in the text. Uncertainties in $\eta_{STS}$, $\varUpsilon_{1,STS}$ and $\alpha_{STS}$ reflect the standard deviation from averaging over $\sigma > 2500$\,Pa.}
\label{fig:shear_thick_phi}
\end{figure}

The relative viscosity $\eta_r(\sigma)$ [Fig.~\ref{fig:rel_visc_n1}(a)] exhibits features characteristic of typical shear thickening colloidal suspensions \cite{Barnes:1989aa, Mewis:2011aa}. At low $\sigma$ there is mild shear thinning, followed by a plateau at a value $\eta_N(\phi)$, which we identify as the high-shear Newtonian plateau (see the Supplemental Materials \cite{sup_note1}). As the stress is further increased, the viscosity begins to increase and then plateaus at a higher value $\eta_{STS}(\phi)$.  The Newtonian plateau viscosity increases with volume fraction as $\eta_N(\phi)=(1-\phi/\phi_0)^{-2}$ with $\phi_0=0.711 \pm 0.007$ [Fig.~\ref{fig:shear_thick_phi}(a)], in good agreement with previous measurements of the high-shear viscosity in hard-sphere colloids \cite{Phan:1996aa, Mewis:2011aa}. We can fit the shear thickened plateau viscosity to the same form $\eta_{STS}(\phi)=(1-\phi/\phi_m)^{-2}$ yielding $\phi_m = 0.592 \pm 0.006$.  Distinct, diverging branches for the Newtonian and shear thickened viscosity plateaus are observed in other systems, though our measured $\phi_m$ is slightly larger than values reported in previous studies\cite{Cwalina:2014aa, Guy:2015aa}. Though both $\eta_N$ and $\eta_{STS}$ increase with $\phi$, the shear thickening onset stress is independent of $\phi$, again consistent with previous experiments \cite{Maranzano:2001aa,Maranzano:2001ab,Cwalina:2014aa,Guy:2015aa}. 
 
To characterize the steepness of the shear thickening, we fit the viscosity increase to $\eta_r \propto \sigma^\beta$. The onset of discontinuous shear thickening (DST) is defined by $\beta=1$, which implies a steady viscosity increases at a fixed shear rate, and $\beta <1$ corresponds to continuous shear thickening. In our suspensions $\beta$ increases monotonically with $\phi$ up to $\beta = 0.95 \pm 0.05$ at $\phi = 0.58$, approaching the DST onset.

While the transition to DST occurs at approximately $\phi =0.58$, $N_1(\sigma, \phi)$ reveals a transition elaborated below that is not evident in $\eta_r$ [Fig.~\ref{fig:rel_visc_n1}(b)] . For $\phi \leq 0.52$, $N_1 \approx 0$ for $\sigma \lesssim 100$ Pa, then drops below zero and becomes increasingly negative as $\sigma$ is increased. The decrease in $N_1$ becomes more pronounced as $\phi$ is increased up to $\phi =0.52$.  At higher volume fractions, $N_1(\sigma)$ initially decreases below zero as before, but as $\sigma$ increases further $N_1(\sigma)$ reverses direction, crosses zero at a shear stress $\sigma_c$ and becomes positive. 

Simulations based on lubrication hydrodynamics predict $N_1<0$ and that $N_1$ should scale linearly with $\dot{\gamma}$ in the high shear limit \cite{Morris:1999aa, FOSS:2000aa}, so that the dimensionless first normal stress coefficient $\varUpsilon_1 \equiv -N_1/\eta_f \dot{\gamma}$ approaches a stress-independent constant. Below $\phi = 0.52$, where $N_1$ remains negative, we find that $\varUpsilon_1$ is indeed stress-independent above $\sigma\approx 100$ Pa, while below this stress we cannot resolve $N_1$ [Fig.~\ref{fig:rel_visc_n1}(c)]. The average value $\varUpsilon_{1,STS}(\phi)$ increases monotonically with $\phi$. An empirical relation 
\begin{equation}
\varUpsilon_{1,STS}(\phi) = K_1 \left( \frac{\phi}{\phi_{max}} \right)^2 \left(1-\frac{\phi}{\phi_{max}} \right)^{-2},
\label{eq:MB}
\end{equation}
initially proposed to capture simulation results \cite{Morris:1999aa}, was shown to fit experimental results for $\phi \leq 0.52$ with $K_1 = 0.177 \pm 0.022$ and $\phi_{max} = \phi_m$ obtained from $\eta_{STS}(\phi)$ \cite{Cwalina:2014aa}. Our results for $\phi\leq 0.52$ can be fit using this same expression [Fig.~\ref{fig:shear_thick_phi}(b)], with a nearly identical coefficient $K_1 = 0.14 \pm 0.01$. 

At higher volume fractions $\varUpsilon_1(\sigma)$ is no longer stress-independent, but instead changes sign as the suspensions become dilatant. As $\phi$ increases, the crossover stress $\sigma_c$ decreases, in contrast to the shear thickening onset stress which remains independent of $\phi$. Below $\sigma_c$,  we can identify a plateau in $\varUpsilon_1$ over a limited range of $\sigma$. This initial plateau follows Eq.\,(\ref{eq:MB}) up to $\phi = 0.56$, even though $\varUpsilon_1(\sigma)$ eventually drops below zero. At $\phi =0.58$, the stress ratio $\alpha = N_1/\sigma$ is approximately constant in the high-stress limit [Fig.~\ref{fig:rel_visc_n1}(d)], consistent with a simple geometric model for force chains \cite{Cates:1998aa}. %At $\phi =0.54$ and $\phi = 0.56$ it is more difficult to discern if $\alpha$ has reached a constant value or continues to slowly increase with increasing $\sigma$. 

To characterize the stress ratio in the high-shear limit, we define $\alpha_{STS}(\phi)$, taking the average over $\sigma \geq 2500$~Pa. Below $\phi=0.52$, where $N_1$ remains negative,  $\alpha_{STS}$ slightly decreases with increasing $\phi$. Noting that $\varUpsilon_1$ can be rewritten as $\varUpsilon_1 = - (N_1/\sigma) \eta_r = -\alpha \eta_r$, we see that the singular term in Eq.\,(\ref{eq:MB}) can be solely ascribed to the viscosity divergence. Thus, as long the stress ratio $\alpha$ remains bounded, fits to Eq.\,(\ref{eq:MB}) are guaranteed to give the same $\phi_{max}=\phi_m$ where $\eta_{STS}$ diverges, but does not imply that $N_1$ and $\eta_r$ are necessarily linked. If lubrication forces drive the rise in $\eta_{STS}$, $N_1$ should become increasingly negative as $\phi \rightarrow \phi_m$. Instead, $N_1$ changes dramatically and becomes positive prior to reaching $\phi_m$, revealing a lack of coupling between the viscosity and $N_1$. %While $\alpha_{STS}$ changes direction and crosses zero between $\phi = 0.52$ and $\phi = 0.54$, there is no obvious change in $\eta_{STS}(\phi)$ [Fig.~\ref{fig:shear_thick_phi}(c)].  Similarly, there is no change in the form of $\eta_r(\sigma)$ that coincides with the change in the sign of $N_1$.

\begin{figure}
\center{\includegraphics[width=7cm]{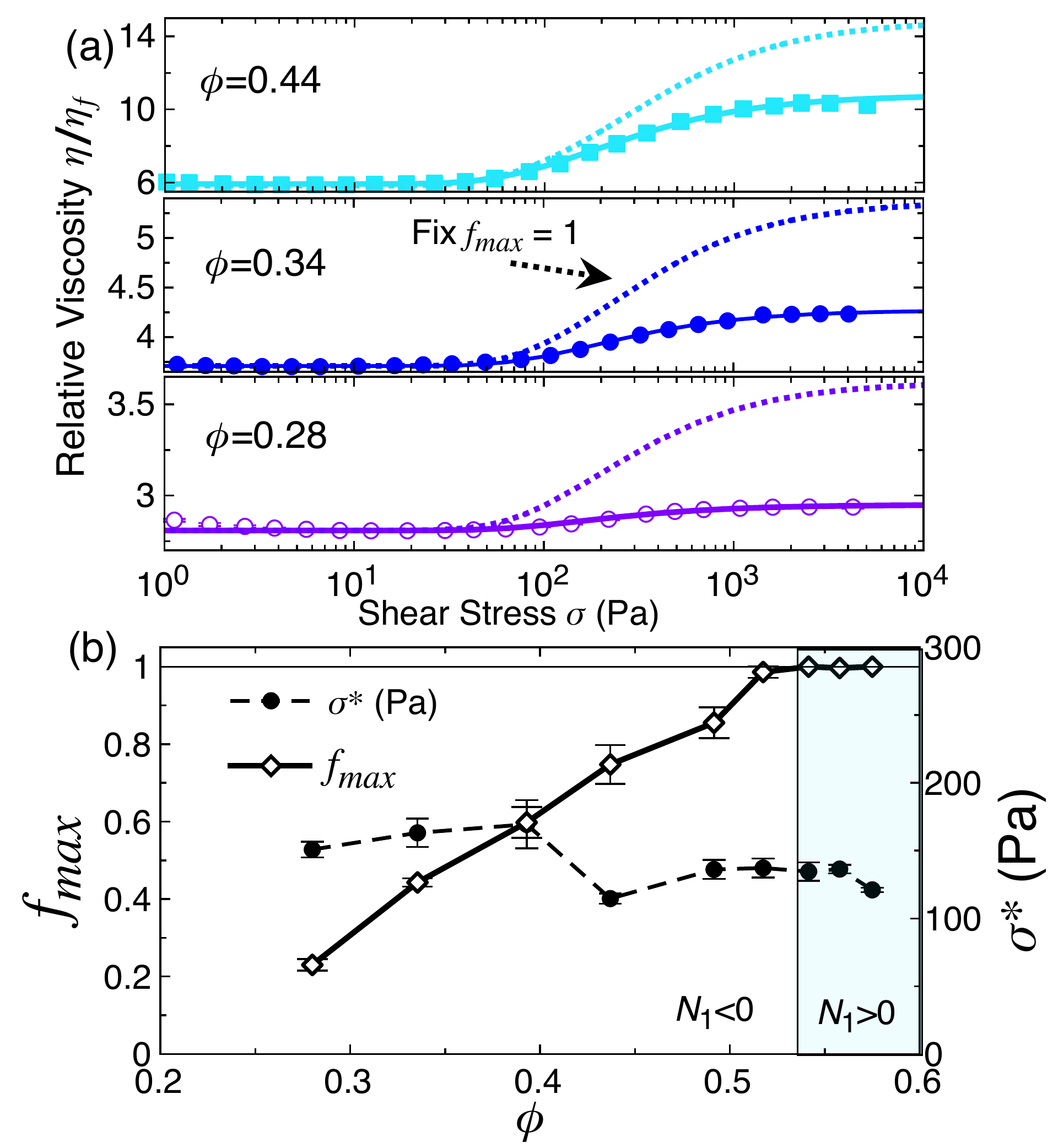}}
\caption{(Color online) Model comparison. (a) $\eta_r(\sigma)$ at selected $\phi$, showing fits to Eq. (\ref{eq:WC}) fixing $f_{max}=1$ (dotted lines) and letting $f_{max}$ vary (solid lines). (b) $f_{max}$ and $\sigma^*$ extracted from fits to Eq. (\ref{eq:WC}). Uncertainties in $\sigma^*$ and $f_{max}$ reflect the standard uncertainty from these fits.   }
\label{fig:wc_model}
\end{figure}  

Positive values of $N_1$ suggest that frictional forces are present and become dominant as $\phi \rightarrow \phi_m$. Motivated by this, we fit $\eta_r(\sigma, \phi)$ using a recently proposed friction-based model \cite{Wyart:2014ve, Guy:2015aa}. The model assumes $\eta_r(\phi)$ is controlled by two distinct divergences, one at $\phi=\phi_0$ for frictionless contacts and a second divergence at a friction-dependent $\phi_m < \phi_0$, giving the two branches $\eta_{STS}$ and $\eta_N$ shown in Fig.~\ref{fig:shear_thick_phi}(a). The full flow curves are given by 
\begin{equation}
\eta_r(\sigma, \phi) = \left( 1-\frac{\phi}{\phi_c(\sigma)} \right)^{-2}
\label{eq:WC}
\end{equation}
where $\phi_c(\sigma) = f \phi_m + (1-f)\phi_0$ interpolates between the two maximum volume fractions and $f\in[0,1]$ represents the fraction of frictional contacts. In this model, contacts become frictional when the compressive force between neighbors exceeds a repulsive stabilizing force $F_{rep}$. While the precise form of $f=f(\sigma,\phi)$ depends on the microstructure and the local contact force distribution, we first adopt a simple ansatz $f(\sigma)=e^{-\sigma^*/\sigma}$. This form is also used in \cite{Guy:2015aa}, which they motivate by assuming an exponential contact force distribution and counting the fraction of local forces above $F_{rep}$, which sets the threshold stress $\sigma^*\propto F_{rep}/d^2$.   

This friction-based model fits our results at high volume fractions, where we find $N_1>0$, exceptionally well [Fig.~\ref{fig:rel_visc_n1}(a)]. Here we hold $\phi_m=0.592$ fixed, but leave both $\sigma^*$ and $\phi_0$ as adjustable parameters. Allowing $\phi_0$ to vary accounts for scatter in $\eta_N(\phi)$, though the fitted values agree with $\phi_0=0.71$ within uncertainty.  At lower volume fractions, where we find $N_1<0$, there is no change in the shear thickening onset nor any qualitative change in the viscosity profile $\eta_r(\sigma)$, suggesting this same model can be applied. Indeed, Figure~\ref{fig:rel_visc_n1} shows that even though $N_1$ is strongly negative for $\phi =0.52$, dropping to as low as -700 Pa, this friction-based model still captures the shear thickening. 

%Since the positive values of $N_1$ strongly suggest the presence of frictional interactions, it is not surprising that shear thickening at high volume fractions can be captured with this friction-based model.

%This agreement is not necessarily unexpected, as negative values of $N_1$ indicate hydrodynamic interactions but do not preclude the presence of frictional contacts. Positive values of $N_1$ arise from system-spanning frictional force chains, 
%We find that $\sigma^*$ is independent of $\phi$, with an average $\langle \sigma^* \rangle =140$ Pa $\pm 18$ Pa. 

The model fits begin to overshoot the amount of shear thickening below $\phi \lesssim 0.5$  [Fig.~\ref{fig:wc_model}(a)]. Though the absolute magnitude of this overshoot is small, this discrepancy can be seen in the limiting viscosities, where the relative difference between measured and fitted values for $\Delta \eta \equiv \eta_{STS} - \eta_N$ increases with decreasing $\phi$  [Fig.~\ref{fig:shear_thick_phi}(a)]. We attribute this disagreement to our simple ansatz for $f(\sigma)$, where $f\rightarrow 1$ for $\sigma \gg \sigma^*$, independent of $\phi$. While we might expect this close to $\phi_m$, in dilute suspensions we expect the flow to be dominated by momentary collisions as opposed to enduring contacts. If we instead take $f(\sigma, \phi) = f_{max}(\phi)e^{-\sigma^*/\sigma}$ with $0 \leq f_{max} \leq 1$, we can fit $\eta(\sigma, \phi)$ over our full range of $\phi$.  We find that the {\it ad hoc} parameter $f_{max} \approx 1$ for $\phi \gtrsim 0.5$, but below this point $f_{max}$ monotonically decreases with decreasing $\phi$ [Fig.~\ref{fig:wc_model}(b)].

To understand the regime where $f_{max}<1$, we posit the formation of enduring frictional contacts requires not only temporary local stresses exceeding $\sigma^*$, but also a confining force to maintain these contacts.At moderate $\phi$ this many-body confinement could be provided by hydrodynamic lubrication forces, reminiscent of the `hydrocluster' model. In this speculative scenario, shear thickening is driven by the formation of frictional contacts within hydrocluster-like structures, though the fraction of frictional contacts would be limited by the size of these clusters so that $f_{max}<1$. Since frictional contacts are confined within these finite clusters, there are no system spanning force chains and the normal stress difference is dominated by lubrication forces, giving $N_1<0$. At some $\phi_{perc}<\phi_m$ these clusters span the system so that $f_{max} \approx 1$ and frictional contact networks percolate throughout the system, driving the transition to $N_1>0$.  %Beyond $\phi_{perc}$ the system is dense enough that hydrodynamic interactions  are not necessary for confinement and the original friction-based model applies.}

The scenario proposed here bridges competing friction-driven and lubrication-driven explanations for colloidal shear thickening. At moderate concentrations, hydrodynamic forces distort the  microstructure and bring particles together, consistent with previous experiments where hydrocluster-like structures have been observed \cite{Cheng:2011aa} and negative values of $N_1$ directly linked to hydrodynamic stresses \cite{Gurnon:2015aa}.  However, the viscosity increase is ultimately driven by the formation of frictional contacts within these clusters, consistent with recent experimental evidence for friction driven shear thickening \cite{Guy:2015aa,Lin:2015aa}. This scenario differs from proposed mechanisms for the onset of DST in granular suspensions \cite{Fall:2008aa, Brown:2012qf}. Instead of dilation driving shear thickening, both dilation and shear thickening are separate consequences of frictional interactions. Dilation requires system-spanning frictional contacts and hence high volume fractions, while shear thickening can result from non-system-spanning frictional contacts and hence occurs over a wider range of volume fractions.   

The transition in the sign of $N_1$ observed here is qualitatively similar to results with roughened silica spheres \cite{Lootens:2005aa}, where the onset of dilation at $\phi= 0.43$ also precedes the transition to DST at $\phi=0.455$. Both transitions occur at lower volume fractions, which we would expect as enhanced roughness should increase the particle friction and decrease $\phi_m$;  which may also explain the difference between values in previous experiments \cite{Cwalina:2014aa, Guy:2015aa}. Recent simulations which include both lubrication and frictional interactions show a similar transition in the sign of $N_1$, with $N_1\lesssim0 $ at $\phi = 0.5$ and $0.53$, but $N_1>0$ at $\phi = 0.55$ \cite{Mari:2015ab}.  Other simulations, which also include lubrication and friction but only explore moderate volume fractions $\phi \leq 0.45$, find that friction weakly increases $N_1$ but overall $N_1$ remains negative \cite{Sierou:2002aa, Gallier:2014aa}, again consistent with our results. Normal stress differences have the potential to serve as a sensitive diagnostic of particle interactions, particularly the presence of frictional interactions. Our results highlight the need for additional studies to determine the effects of particle size, roughness and other surface properties. 

\begin{acknowledgments}
We thank J. Seppala and E. Del Gado for insightful discussions and A. Forster for assistance with supplementary tests. D.L.B. was supported by US National Science Foundation Grant No. DMR-0847490 and Department of Commerce Cooperative Agreement 70NANB15H229. Official contribution of the National Institute of Standards and Technology; not subject to copyright in the United States. 
\end{acknowledgments}

\bibliography{../../../../Dynamics_Cubic_Colloids/cubic_colloids.bib}

\begin{thebibliography}{39}
\bibitem[{nis()}]{nist_prod}
\bibinfo{note}{Certain instruments and materials are identified in this paper
  to adequately specify the experimental details. Such identification does not
  imply recommendation by the National Institute of Standards and Technology
  nor does it imply the materials are necessarily the best available for the
  purpose.}
\end{thebibliography}
\bibliographystyle{apsrev}

\end{document}